\documentclass[final]{svjour2}
\usepackage{graphicx}
\usepackage{rotating}
\usepackage{amssymb}
\usepackage{mathptmx}
\usepackage[numbers]{natbib}
\makeatletter \journalname{Journal of Low Temperature Physics}

\bibpunct{}{}{,}{s}{}{,}

\begin{document}

\newcommand{\hdblarrow}{H\makebox[0.9ex][l]{$\downdownarrows$}-}
\title{Topological matter: graphene and superfluid  $^3$He}

\author{M.I. Katsnelson$^{1,2}$ and  G.E. Volovik$^{3,4}$
}
\institute{
1:
Radboud University Nijmegen, Institute for Molecules
and Materials, Heyndaalseweg 135, NL-6525AJ Nijmegen, The Netherlands
\\
2: Department of Theoretical Physics and Applied Mathematics, Ural Federal University, 620002 Ekaterinburg, Russia
\\ Tel.: +31-24-3652995, Fax: +31-24-3652120, \email{M.Katsnelson@science.ru.nl}
\\
3: Low Temperature Laboratory, Aalto University, P.O.
Box 15100, FI-00076 AALTO, Finland
\\ 4: L.D. Landau Institute for
Theoretical Physics, 119334
 Moscow, Russia
 \\ Tel.: +358-9-4512963, Fax: +358-9-4512969, \email{volovik@boojum.hut.fi}
 }

\date{\today}

\maketitle


\begin{abstract}
The physics of graphene and of the superfluid phases of $^3$He have many common features. Both systems are topological materials where quasiparticles behave as relativistic massless (Weyl, Majorana or Dirac) fermions. We formulate the points where these features are overlapping. This will allow us to use graphene to study the properties of superfluid $^3$He, to use superfluid $^3$He to study the properties of graphene, and to use both of them in combination to study the physics of topological quantum vacuum. We suggest also some particular experiments with superfluid $^3$He using graphene as an atomically thin membrane impenetrable for He atoms but allowing for spin, momentum and energy transfer.

\end{abstract}

\section{Introduction}
\label{Introduction}

Topological media are gapped or gapless fermionic systems, whose properties are protected by topology, and thus are robust to deformations of the parameters of the system and generic\cite{VolovikBook2003,Katsbook,Qi2011}.

Among the gapless topological media (topological vacua) one may find normal metals, chiral Weyl superfluid $^3$He-A, graphene, and also the relativistic quantum vacuum of Standard Model in its symmetric phase, which represents the class of Weyl semimetals. These media have topologically protected zeroes in the energy spectrum of fermionic (quasi)particles, which are characterized by topological invariants. The nodes in the spectrum may form Fermi surfaces (in normal metals), Weyl points (in $^3$He-A and relativistic vacuum), and Dirac points (in graphene and at the surface of three-dimensional topological insulators). The vacua with Weyl and Dirac points serve as a source of effective relativistic quantum field theories (QFT) emerging at low energy. In the vicinity of the Weyl point, left-handed and right-handed relativistic fermions emerge, which  live in the effective 3+1 dimensional spacetime, described by the effective tetrad gravity with spin connection and torsion, and interact with effective gauge fields. Accordingly, the Dirac fermions in graphene  live in the effective 2+1 dimensional spacetime. The accompanying effects (chiral and gravitational anomalies; chiral magnetic and chiral vortical effects; electroweak baryo-production and its observed analogue in superfluid $^3$He; etc.) are expressed via symmetry protected topological invariants in terms of the Green's function. Systems with degenerate  Dirac points, such as bilayer and multilayer graphene, may have Dirac fermions with nonlinear spectrum (quadratic, cubic, quartic, etc.) and experience emergent QFT with anisotropic scaling and the Ho\v{r}ava-Lifshitz gravity (see Refs.\cite{KVZ1,KVZ2} and references therein).

According to the bulk-surface and bulk-vortex correspondence, the gapless topological media have exotic fermions living on the surface of the system or within the core of topological defects, where they form a Fermi arc or a flat band. A flat band appears in particular on the zigzag edges of graphene (assuming exact electron-hole symmetry, e.g., in the nearest-neighbor approximation): all electrons within the flat band have exactly zero energy\cite{Katsbook}. At some conditions the flat band may emerge in the multilayered graphene, where it forms a kind of graphite with rhombohedral (ABC) stacking.  The existence of the flat band crucially influences the critical temperature of the superconducting transition. While in all the known superconductors the transition temperature is exponentially suppressed as a function of the pairing interaction, in the flat band the transition temperature is proportional to the pairing interaction, and thus can be essentially higher\cite{HKV2011}. The search for or artificial fabrication of such semimetals may therefore open a route to room-temperature superconductivity. Alternative channels of instability, such as magnetic instability, should be also taken into account. This all makes systems with flat bands a very interesting and promising field.

The topological media, which are gapped in the bulk, have topologically protected surface excitations, such as Majorana fermions\cite{Wilczek2009,Mourik2012}. This class of topological media includes topological insulators, superfluid $^3$He-B, thin film of $^3$He-A, and the quantum vacuum of the Standard Model in the broken symmetry phase.  The vacua of this class exhibit quantization of physical parameters expressed via symmetry protected topological invariants. Graphene may belong to this class if a broad enough energy gap appears due to spin-orbit interaction (or due to spontaneously broken symmetry), see, e.g., Ref.\cite{Kou2013}. Then as a 2+1-dimensional topological medium,  it may experience, say, the anomalous quantized Hall and spin Hall effects without external magnetic field\cite{Qi2011}.

Here we discuss a program of future studies of those properties which are common to both superfluid $^3$He and graphene and experiments with graphene in the superfluid $^3$He environment. Both systems are topological materials. They contain topologically protected massless fermions: 2+1 Dirac fermions in graphene\cite{Katsbook}; 3+1 Weyl fermions in bulk $^3$He-A; 2+1 Majorana fermions on the surface of $^3$He-B; 1+1 Majorana fermions in the cores of quantized $^3$He-B vortices\cite{VolovikBook2003}. In both systems relativistic quantum fields and gravity emerge with all the related phenomena, such as the chiral anomaly, Hawking-Unruh effects, and Schwinger pair production in electric field\cite{KVZ1,KVZ2}. The combination of graphene and superfluid $^3$He makes it possible to study the interplay of the properties of two topological materials plus the new effects, which emerge, when these materials are combined.

This program contains the following issues. In Sec. 2 we consider the new symmetry breaking phenomenon in superfluid $^3$He, which is possible due to graphene membrane: formation of exciton condensate between  $^3$He atoms located on the two sides of the ``brane''. In Sec. 3 the ripplons on a graphene membrane immersed in superfluid $^3$He are considered. Sec. 4 discusses  Dirac fermions in graphene and Majorana fermions on the two surfaces of $^3$He-B. The ripplons on the brane  serve as common gravitational field acting on these ``relativistic'' fermions living on three different branes. Sec. 4 is devoted to spin current Josephson effect between two magnon Bose condensates in $^3$He-B separated by the graphene membrane, which can serve as a sensitive tool to study  the ``relativistic'' fermions and also to study the possible magnetic states of graphene discussed in Sec. 5.

\section{Exciton condensate across graphene}
\label{Exciton}

One direction to probe the topologically protected  ``relativistic'' fermions in the combined system is to consider the transport of mass and spin currents of superfluid $^3$He through graphene. The graphene is thus used as the interacting atomic-size interface between identical or different superfluid phases of $^3$He and as the mass current and spin current Josephson junctions. In this paper we consider only the spin currents, since in electrically neutral superfluid $^3$He the NMR technique is the most sensitive tool.
So we discuss the situation when there are no holes in the graphene membrane and thus there is no penetration of $^3$He atoms through the graphene sheet, since the potential barrier is more than 10 eV and therefore the tunnelling time is astronomically long (for permeation of helium through graphene see Refs.\cite{Bunch2008,LPP2008,Drahushuk2012}). In this case the $^3$He atoms on the two sides of graphene can be considered as two independent fermionic species. However, energy, momentum and spin transfer is possible. This in particular leads to possibility to obtain the exciton condensate in superfluid $^3$He, which we discuss in this section.

To simplify the problem, let us neglect the orbital and spin degrees of freedom of the $^3$He order parameter and concentrate only on the gauge group $G=U(1)$, whose spontaneous breaking gives the phenomenon of superfluidity. If the tunneling of atoms through graphene is absent, the global gauge symmetry of two species in the normal state is
$G\times G=U(1) \times U(1)$. When both species are in the superfluid state,  this group is broken to
the subgroup $H\times H =Z_2\times Z_2$, where the $Z_2$-groups  are gauge rotations by $\pi$ for each species due to the formation of two condensates:
\begin{equation}
\left<\Psi_1\Psi_1\right> =A_1e^{i \Phi_1} ~~,~~\left<\Psi_2\Psi_2\right> =A_2e^{i \Phi_2}   \,.
\label{separate}
\end{equation}
Here indices 1 and 2 refer to $^3$He atoms on the two sides of graphene. Due to interaction across graphene and via graphene, say, due to ripplon exchange, or even due to direct van der Waals interaction of helium atoms through a thin membrane, the $Z_2\times Z_2$ group can be further broken to  the $Z_2$ subgroup by the formation of cross correlations in the Cooper channel
\begin{equation}
\left<\Psi_1\Psi_2\right> =Be^{i \phi}  \,,
\label{cross-correlator1}
\end{equation}
and related correlations in the particle-hole channel (exciton condensate):
\begin{equation}
\left<\Psi_1\Psi_2^\dagger\right> =Ce^{i \varphi}  \,.
\label{cross-correlator2}
\end{equation}
These correlations are not independent. If, say, the pairing occurs in the Cooper channel in Eq.(\ref{cross-correlator1}), then the correlator $\left<\Psi_1\Psi_2^\dagger\right>$ will  also appear  as the subdominant term due to symmetry reason, and vice versa.
The phases $\varphi$ and $\phi$ can be expressed via the bulk phases
$\Phi_1$ and $\Phi_2$
by minimization of the ``Josephson couplings'':
\begin{equation}
\lambda_1\cos(\Phi_1 +\Phi_2 - 2\phi)  +\lambda_2\cos(\Phi_1-\Phi_2 - 2\varphi)
+\lambda_3\cos(\Phi_1 - \phi-\varphi) + \lambda_4\cos(\Phi_2 - \phi+\varphi)
\label{phase_couplings}
\end{equation}
The phase  transition $Z_2\times Z_2 \rightarrow Z_2$ can be described by a discrete Ising order parameter. If the dominant mechanism is  the  exciton condensation in Eq.(\ref{cross-correlator2}), the order parameter is
 \begin{equation}
M=\frac{\left<\Psi_1\Psi_2^\dagger\right>}{\left<\Psi_1\Psi_1\right>^{1/2} \left<\Psi_2^\dagger\Psi_2^\dagger\right>^{1/2}}  \,.
\label{Ising1}
\end{equation}
The order parameter $M$ changes sign under gauge transformation by $\pi$ for any of the two species, but does not change
sign under the combined gauge transformation by $\pi$ of both species. The latter combined symmetry is the rest symmetry $Z_2$ of the system.
If the dominant mechanism is the Cooper pairing in Eq. (\ref{cross-correlator1}), the order parameter
will be
\begin{equation}
M=\frac{\left<\Psi_1\Psi_2\right>}{\left<\Psi_1\Psi_1\right>^{1/2} \left<\Psi_2\Psi_2\right>^{1/2}}  \,.
\label{Ising2}
\end{equation}
The system may have domain walls (lines) on the graphene, across which the order parameter $M$ changes sign.

Here we considered only the gauge $U(1)$ part of the symmetry group $G$. In general the exciton condensate may contain spin and orbital degrees of freedom, since the symmetry group of two species of fermions is
$G\times G=[U(1)\times SO(3)_S \times SO(3)_L]\times [U(1)\times SO(3)_S \times SO(3)_L]$, which in $^3$He-B is broken to the subgroup $H\times H=[U(1)\times SO(3)_J]\times [U(1)\times SO(3)_J]$. This
 would result in different classes of broken symmetry, which influence the  boundary conditions and  result in different types of the topological defects on the graphene sheet (such as boojums), and different rules for the termination of vortices and strings on the graphene  (see Sec. 17.3 in Ref.\cite{VolovikBook2003}).
In the anisotropic $^3$He-A the connection between the orbital $\hat {\bf l}$ and spin $\hat {\bf d}$ anisotropy axes across the graphene sheet takes place even without the additional symmetry breaking: there are the terms in the surface energy
$a_{12} \hat {\bf l}_1 \cdot \hat {\bf l}_2$ and $b_{12}  (\hat {\bf d}_1 \cdot \hat {\bf d}_2)^2$.

The open questions left for future consideration: what is the transition temperature  of this exciton condensation; what type of condensate is dominating; what kind of fermions live on the domain line; what happens with the tangential superflow across graphene; what if graphene serves as the interface between $^3$He-B and $^3$He-A; how the possible solid layer of $^3$He on graphene influences
the exciton condensation; etc.

Another interesting opportunity is to use a graphene membrane with holes; according to  computations\cite{LPP2008} the height of the barrier for the He atom drops to a value of the order of 1 eV for tetravacancy and of the order of 0.1 eV for decavacancy. Cutting holes in graphene one can prepare a true Josephson junction.

Numerical simulations of helium atoms on graphene are preformed in Ref.\cite{Kwon2012}, and on graphane and graphene-fluoride in Ref.\cite{Reatto2012}. Keeping in mind also the possibility of solid helium formation at the membrane, a lot of additional work should be done before it will be possible to discuss the opportunity suggested here quantitatively. However, conceptually it looks interesting and for sure deserves further studies.

\section{Ripplons on graphene membrane}
\label{Ripplons}

\subsection{Graphene sheet immersed in bulk liquid}

Oscillations of the graphene membrane are important, e.g., as a source of an effective interaction between helium atoms in the vicinity of the membrane. The standard theory of membrane oscillations (see Ref.\cite{Katsbook}, Chap.9) is not applicable for a thin membrane in liquid. In the latter case, the membrane mass is renormalized by the associated hydrodynamic mass of superfluid components in $^3$He-B and $^3$He-A. The hydrodynamics leads to a modification of the $\omega \propto k^2$ ($\omega$ is the frequency and $k$ is the wave vector) spectrum of ripplons (that is, acoustic flexural phonons)\cite{Katsbook} for freely suspended graphene.

The long-wavelength oscillations of the graphene sheet immersed into superfluid $^3$He
\begin{equation}
 h(x,t)=a\sin (kx-\omega t) \,,
\label{zeta}
\end{equation}
have the following spectrum (assuming the same superfluid state across the graphene):
\begin{equation}
 \rho_g \omega^2+ \frac{\rho}{k}\left({\omega}-kv_1\right)^2+ \frac{\rho}{k}\left({\omega}-kv_2\right)^2
=\sigma k^2 +\kappa k^4-i\Gamma \omega \,.
\label{Spectrum}
\end{equation}
It can be obtained from Eq. (27.8) of Ref. \cite{VolovikBook2003} for the interface between two superfluids by adding the mass and rigidity of graphene.
Here $\rho_g$ is the 2-dimensional mass density of graphene; in liquids  the associated hydrodynamical 2-dimensional mass density $\rho/k$ is added due to dynamics of bulk liquid, where $\rho$ is the 3-dimensional mass density of liquid $^3$He; $\kappa$ is the bending rigidity of graphene;
$\sigma$ is the surface tension; the gravitational force is absent since the density $\rho$ of liquid $^3$He is assumed to be the same on two sides of the sheet;  $\Gamma$ is the friction parameter;  ${\bf v}_1$ and ${\bf v}_2$ are the velocities of superfluid $^3$He on the two sides of the sheet, which give rise to the Doppler shifts in the spectrum. We assume here that both superfluid velocities are along the $x$ axis
(${\bf v}_1=v_1\hat{\bf x}$ and ${\bf v}_2=v_2\hat{\bf x}$).

Since the gravity or its magnetic field counterpart are absent, both the Kelvin-Helmholtz instability for $v_1\neq v_2$ and the ergoregion instability \cite{VolovikBook2003} for $v_1= v_2$ have no threshold. This means that the instability should develop at any rotation velocity, but it is limited by the finite size effect.
The first term on the right-hand side of Eq. (\ref{Spectrum}) is compared with the second one at
$k_c \approx \sqrt{\sigma/ \kappa} \approx 4 \cdot 10^5$ cm$^{-1}$ where we
used the values $ \sigma \approx 0.1$ erg/cm$^2$, $\kappa \approx 6 \cdot 10^{-13}$ erg.
This means that for any reasonable sizes of the membrane $L$ one has $Lk_c \gg 1$, so that
the bending rigidity of graphene is negligible, and the ripplon spectrum is similar to that at the liquid interfaces.

The modification of the ripplon spectrum of graphene in superfluid $^3$He as compared with  graphene suspended in the vacuum may influence the transport properties of graphene, if the resistance is determined by interactions of electrons with flexural phonons (ripplons)\cite{Castro2010,Ochoa2011,Gornyi2012}.

The open questions left for future consideration: what is the contribution of topologically protected Majorana fermions on the surface of $^3$He-B to the frequency shift and frictional dissipation; what is the contribution of orbital viscosity in bulk $^3$He-A which comes from the topologically protected Weyl fermions;  how do the Majorana and Weyl fermions modify the spectrum of short-wavelength oscillations; what is the effect of the solid layer of $^3$He, if it is formed on the graphene sheet;
what happens if the graphene separates A and B phases in an applied gradient of the magnetic field, which is equivalent to a gravitational field; etc.

The turbulence of gravity waves  on the surface of $^3$He-B is a possible source of the observed extra dissipation of surface oscillations\cite{Eltsov2013}.  The graphene sheet may influence the superfluid wave turbulence. An analogous effect has been discussed in normal liquid coated by an elastic sheet\cite{Deike2013}, a related phenomenon is the smoothing effect of an oil film on the surface of the ocean.

\subsection{Graphene sheet in a slab and shallow-water limit}

The more general expression for the ripplon spectrum in a slab is
\begin{equation}
M_1(k) (\omega- {\bf k}\cdot {\bf v}_1)^2
+ M_2(k)(\omega- {\bf k}\cdot {\bf v}_2)^2+\rho_g \omega^2
=  F+k^2\sigma +\kappa k^4-   i \Gamma \omega~.
\label{GeneralRipplonSpectrum}
\end{equation}
Here $F$ is the applied
gradient of magnetic field; $h_1$ and $h_2$ are the thicknesses of the layers of two
superfluids;
$M_1(k)$ and
$M_2(k)$ are the $k$-dependent masses of the liquids which are
forced into motion by the oscillating brane:
\begin{equation}
M_1(k)=\frac{\rho}{k ~{\rm tanh}~kh_1}~,~M_2(k)=\frac{\rho}{k~ {\rm tanh}~kh_2} \,.
\label{Masses}
\end{equation}
The spectrum of ripplons becomes relativistic in the shallow-water limit $kh_1\ll 1$, $kh_2 \ll 1$. For ${\bf v}_1={\bf v}_2=0$ one has $\omega=ck$ with $c^2=(F/\rho) (h_1 h_2)/(h_1+h_2)$, while the velocity fields ${\bf v}_1$ and ${\bf v}_2$ give rise to the effective metric acting on these 2-dimenional bosonic ``relativistic'' particles \cite{VolovikBook2003}. In particular, this will allow us to use graphene to study the effects of event horizons for ripplons.
Also, by changing the thicknesses $h_1$ and $h_2$ of the superfluid layers one may influence the graphene conductivity.

\section{Weyl, Dirac and Majorana fermions}
\label{Fermions}

There are several different types of ``relativistic'' fermions living in the system.

First there are  3+1 dimensional chiral Weyl fermions living in bulk superfluid $^3$He-A. The gapless spectrum of a
single Weyl fermion species is protected by topology and cannot be destroyed by any perturbation
\cite{VolovikBook2003}.
The Weyl point disappears only when it merges with the Weyl point of opposite chirality, which
has the opposite topological charge. This does not occur in $^3$He-A, where the Weyl point with left fermions and the Weyl point with right fermions are well separated in momentum space, but the Weyl points may merge in a gas of cold
fermionic atoms with $p$-wave pairing.   Then the hybridization of two species (left-handed and right-handed fermions) gives rise to the 3+1 Dirac fermion.
The Dirac fermions remain gapless if there is some special symmetry
which protects the gaplessness. This is the symmetry which supports the topology,
i.e. there exists the topological invariant which is nonzero only in the
presence of the symmetry.  When this symmetry is violated or spontaneously
broken, as happens in the Standard Model, then there is no more topological protection and the Dirac fermion acquires mass.

The system of graphene immersed in $^3$He-B contains 2+1 dimensional ``relativistic'' fermions. These are four species of  fermions of graphene (two valleys and two spin projections) and two species of  fermion zero modes living on two boundaries of $^3$He-B  (edge states).

Electrons in graphene  are 2+1 Dirac fermions. As in the case of 3+1 Dirac fermions, their gaplessness is protected by  symmetry, which supports the topology. When symmetry is
violated (by magnetic field or spin-orbit interaction) the mass (gap) may
appear. Massless Dirac fermions in graphene have the spectrum $E=v_F p$ with $v_F \approx 10^8$ cm/s.

If one ignores the superfluidity of $^3$He-B, then it represents the first known example of a  ``topological insulator''. Its fermionic edge zero modes serve as the prototype of the topologically protected gapless fermions on the boundary of the solid-state topological insulators \cite{Qi2011} (topological invariant for the time-reversal invariant topological phase of superfluid $^3$He has been introduced in \cite{SalomaaVolovik1988}, where also the fermionic edge modes at the interface between states of
$^3$He-B with different orientations of the order parameter have been discussed). However, since  $^3$He-B is a Cooper paired superfluid and its fermions are hybrids of particles  and holes, the edge modes represent the Majorana fermions.
The effective Hamiltonian for the Majorana mode on a single boundary of $^3$He-B is\cite{Volovik2009,Volovik2010}
 \begin{equation}
H_{\rm zm}({\bf k})= c_B \hat{\bf n} \cdot({\mbox{\boldmath$\sigma$}} \times {\bf p}) \,,
\label{eq:ModesH}
\end{equation}
where $ \hat{\bf n} $ is the direction of the normal to the graphene sheet;  ${\mbox{\boldmath$\sigma$}}$ are Pauli matrices of spin rotated by the $^3$He-B order parameter matrix $R_{\alpha i}$; the characteristic ``speed of light'' of Majorana fermons $c_B\approx \Delta/p_F \approx $ 1 cm/s is eight orders of magnitude smaller than that of Dirac fermions in graphene.

The relativistic form of the energy spectrum of Dirac and Majorana fermions allows us to use both fermionic species for the simulation of relativistic quantum fields and gravity. In particular, ripplons (flexural phonons) serve as the gravity field for both excitations. Though  Dirac and Majorana fermions  live in different worlds, they interact through the common gravitational field -- the effective zweibein field and correspondingly the effective metric field $\delta g_{ij} \propto \partial_ i h ~ \partial_ j h$ produced by ripplons\cite{Manes-de Juan-Sturla-Vozmediano2013,Zubkov-Volovik2013}. Actually we have
three flavors of ``relativistic'' fermions, which live on three branes (graphene and two surfaces of superfluid $^3$He-B on the two sides of graphene).

The Dirac mass in the undoped graphene comes from the spin-orbit interaction and is small since the coupling is weak (see Ref.\cite{Katsbook}, Sect.12.4). It would be interesting to study the influence of the broken time reversal symmetry in $^3$He-A on the Dirac mass in graphene. The interaction may depend on the orientation of the orbital $\hat{\bf l}$-vectors on the  two sides of graphene, and on the magnitude and orientation of the magnetic field\cite{MizushimaSatoMachida2012}. The possible influence of the spin degrees of freedom of $^3$He-B on the Dirac mass should also be considered. All this may lead to different classes of topological insulator states of graphene characterized by the symmetry protected Chern numbers\cite{VolovikBook2003}.

Majorana fermions of $^3$He-B localized on the two sides of graphene may also acquire mass, say,  due to interactions across the graphene. The possible channels of interaction between the Majoranas are: spin-spin interaction\cite{ChungZhang2009}; interaction via ``gravitational waves'' -- ripples of graphene; etc. Also the effect of the exciton condensation must be considered.

The mass of Majorana fermions may also appear due to the interaction of their spins  with magnetic impurities localized on graphene, as considered for surface states in topological insulators\cite{Shou-Cheng_Zhang2009}. If the magnetic moments of the impurities are in the graphene plane, then there is a threshold field at which the mass appears \cite{MizushimaSatoMachida2012}. In this case the graphene ripples will provide the mass. All this may lead to different classes of the gapped 2+1 topological states on the boundaries of $^3$He-B (see e.g. \cite{Volovik2010}), which may result in different types of the quantized spin Hall conductivity along the brane.

One should also consider the possibility, that the  hybridization of Majorana fermions across graphene  may lead to fermions with quadratic Dirac spectrum, as happens in a bilayer graphene. If this occurs one would have a new example of emergent QFT with anisotropic scaling and the Ho\v{r}ava-Lifshitz gravity.

\section{Spin current physics}
\label{SpinCurrent}

In superfluid $^3$He-B the phenomena related to spin currents have been studied experimentally for several decades, while in solid state physics spintronics is a relatively new and rapidly developing area.  One may expect that the study of the combined spintronics in the system $^3$He-B + graphene will open a new chapter in spin current physics. The most pronounced event in the $^3$He-B  spintronics
was the discovery in 1984 of spontaneous phase coherent spin precession\cite{HPDexp1984,HPDexp1985,Fomin1984,Fomin1985}. It represents spin superfluidity, characterized by off-diagonal long-range order, and can be described in the language of the Bose-Einstein condensation of magnons
(see review\cite{BunkovVolovik2013}). A magnon Bose condensate in a controlled magnonic trap (which is analogous to the so called $Q$-ball in particle physics) serves now as a tool to study different spin and orbital phenomena in  $^3$He-B with great precision, see e.g. recent experiments  \cite{Autti2012a,Autti2012b,Heikkinen2013}.
That is why these methods could be sensitive enough for studies of spin currents in graphene  and through graphene, including the quantized spin-Hall effect\cite{VolovikYakovenko1989}.

One of the directions of further investigations is to study the spin current Josephson effect through graphene by the creation of two droplets of magnon BEC on the two sides of the graphene.
The difference in chemical potentials of the two Bose condensates produces oscillations with period $T=2\pi\hbar/|\mu_2-\mu_1|$, if there is a spin current between the condensates (the first observation of the dc and ac spin current Josephson effect between two magnon condensates connected by narrow channel has been reported in Ref.
\cite{JosephsonEffect1988}).
There are several methods to create the difference in chemical potential in a controlled way: by the difference in the well potentials, by changing the distances between the wells and graphene, by population imbalance, etc. Another possibility to study the spin current through graphene is to pump magnons in one of the traps and detect the formation of magnon BEC in the other trap.

There are different channels of the Josephson coupling of the spin degrees of freedom across graphene: electronic subsystem of graphene; magnetic layer of solid $^3$He; magnons; fermionic quasiparticles in bulk and on the surface (Majorana fermions); dynamics of graphene sheet (ripplons); direct dipole interaction of spins of $^3$He atoms across graphene; propagation of quasiparticles through the graphene (which will also contribute to the heat current).

\section{Low dimensional magnetism of graphene}
\label{Magnetism}
Magnon condensates can be used also for the study of the graphene magnetic states, which may spontaneously emerge at the sub-millikelvin temperatures of superfluid $^3$He.

Experiments\cite{Nair2013} demonstrate the existence of localized magnetic moments associated with vacancies and some adatoms and admolecules ($sp^3$ impurities), of the order of $\mu \approx 1 \mu_B$. At temperatures $T > 1$ K these moments can be considered as noninteracting, with their paramagnetic susceptibility following the Curie law, $\chi \propto 1/T$. One can assume, however, that at low temperatures they should interact and can therefore be ordered by the interaction. Indeed, a typical energy of dipole-dipole interaction is
\begin{equation}
E_{dd} \approx \frac{\mu_B^2}{R^3}
\label{dipdip}
\end{equation}
where $R$ is a typical distance between the magnetic defects. Assuming $R \approx 1$ nm (which correspond to the defect concentration about $10^{-4}$ per atom) we have $E_{dd} \approx 1$ mK which gives us an estimate for the ordering temperature even if we fully neglect the indirect (RKKY) exchange interactions. The latter are, however, important (for review, see Ref.\cite{Kotov2012}). For undoped graphene, their typical energy is
\begin{equation}
E_{dd} \approx \frac{I^2}{tR^3}
\label{exchange}
\end{equation}
where $I$ is the s-d exchange coupling constant and $t \approx 3$ eV is the nearest-neighbor hopping parameter in graphene; this interaction is ferromagnetic for defects in the same sublattice and antiferromagnetic for different sublattices\cite{Kogan2011}. For the doped graphene with a finite Fermi wave vector $k_F$ it oscillates and decays at $R \gg 1/k_F$ as $1/R^2$. Anyway, it is always stronger than the dipole-dipole interaction. The combination of the exchange and dipole-dipole interactions can lead to magnetic ordering even in the two-dimensional case, despite the Mermin-Wagner theorem; this is due to a long-range character of the dipole-dipole interactions\cite{Maleev1976,Grechnev2005}.

NMR of  $^3$He can be used to study this interesting case; recently, it was used to study two-dimensional ferromagnetism of $^3$He adsorbed on the graphite surface\cite{Casey2013}.

\section{Conclusion}
\label{Conclusion}

The superfluid phases of $^3$He and graphene remain at the moment the best and the cleanest representatives of topological condensed matter systems in 3+1 and 2+1 dimensions correspondingly.
Their hybridization opens a new channel in investigations of topological materials and emergent relativistic quantum field theories.  The combination of graphene and superfluid phases of $^3$He can be considered as  an extended topological vacuum. The fermionic excitations of this vacuum represent different species of interacting ``relativistic'' fermions living either on three branes (graphene + two surfaces of $^3$He-B) or on the branes and in bulk (graphene + surface of $^3$He-B + bulk $^3$He-A). The fermions have large difference in the effective ``speed of light'', but they interact via the common effective gravitational field.  For this combined system, we expect new types of the topological order, which are accompanied by the broken or emergent symmetry and can be probed by NMR of $^3$He.

\section*{Acknowledgements}

We acknowledge financial support by the EU 7th Framework Programme (FP7/2007-2013, grant $\#$228464 Microkelvin), GEV by the Academy of Finland through its LTQ CoE grant (project $\#$250280) and MIK by the Nederlandse Organisatie voor Wetenschappelijk Onderzoek (NWO) via Spinoza Prize.

 \end{document}